\documentclass[nofootinbib, reprint,amsmath,amssymb,aps,prl]{revtex4-1}

\usepackage{graphicx}

\usepackage{dcolumn}
\usepackage{bm}
\usepackage{caption}
\usepackage{subcaption}
\usepackage[labelfont=bf,textfont=normalfont,singlelinecheck=off,labelformat=simple]{subcaption}
\usepackage[labelfont=bf,textfont={sf,small},labelsep=none,justification=raggedright]{caption}

\usepackage{hyperref}

\usepackage{enumerate}
\usepackage{float}
\usepackage{xargs}
\usepackage{setspace}
\usepackage[export]{adjustbox}
\usepackage{color}

\newcommand{\blau}[1] {{ {#1}}}

\begin{document}

\title{Direct observation of swap cooling in atom-ion collisions}

\author{Amir Mahdian, Artjom Kr\"ukow  \& Johannes Hecker Denschlag}
\affiliation{Institut f\"ur Quantenmaterie and Center for Integrated Quantum Science and Technology IQST, Universit\"at Ulm, 89069 Ulm, Germany.}

\date{\today}


\begin{abstract}
{
Collisions with cold particles can dissipate a hot particle's energy and therefore can be exploited as a cooling mechanism. Kinetics teach us that cooling a particle down by several orders of magnitude typically takes many elastic collisions as each one only carries away a fraction of the collision energy.
Recently, for a system comprising hot ions and cold atoms, a very fast cooling process has been suggested\cite{1} where cooling over several orders of magnitude can occur in a single step. Namely, in a homo-nuclear atom-ion collision, an electron can resonantly hop from an ultracold atom onto the hot ion, converting the cold atom into a cold ion. Here, we demonstrate such swap cooling in a direct way as we experimentally observe how a single energetic ion loses energy in a cold atom cloud.
In order to contrast swap cooling with sympathetic cooling, we perform the same measurements with a hetero-nuclear atom-ion system, for which swap cooling cannot take place, and indeed observe very different cooling dynamics. Ab initio numerical model calculations agree well with our measured data and corroborate our interpretations.}\\
\end{abstract}

\maketitle
\section{Introduction}
\indent The preparation of cold ions is often a precondition for modern experiments in various scientific fields, ranging from ultracold chemistry\cite{3} to quantum information processing\cite{4}. While laser cooling has opened up research in the ultracold regime\cite{5}, it is generally limited to the species with closed cycling transitions.
Another  method commonly used for cooling  particles to sub-kelvin temperatures is collision-induced cooling, such as sympathetic cooling or buffer gas cooling \cite{3,6,7,8,9,14}.
In addition to translational degrees of freedom, collisions can also cool the molecule's internal
degrees of  freedom  \cite{10,11}.
Sympathetic cooling of a highly energetic particle to orders of magnitude lower kinetic energies
typically requires many collisions, because on average, each collision can only cool away a  small fraction of the energy. Swap cooling, by contrast, can turn a hot ion into a cold one in a single collision\cite{1}. In order to lay out the swap cooling process, we consider a homo-nuclear atom-ion system, \blau{which consists of two identical ion cores and one valence electron. }
  As the hot ion passes by a cold neutral atom at close proximity, the electron can resonantly hop from the atom onto the ion, \blau{as the
the binding energy of the electron is independent of which ionic core it is bound to.
During this process momentum exchange between atom and ion is negligible, (see Fig. 1). }
Thus, the former neutral atom has been converted into a cold ion. First indirect experimental evidence for swap cooling was recently observed in terms of an increased lifetime of Cs$^+$ ions in an ion trap  when sympathetically cooled by an ultracold cloud of Cs atoms instead of Rb atoms\cite{1,2}.

\begin{figure}[ht!]
	\centering
	\includegraphics[width=8.2cm]{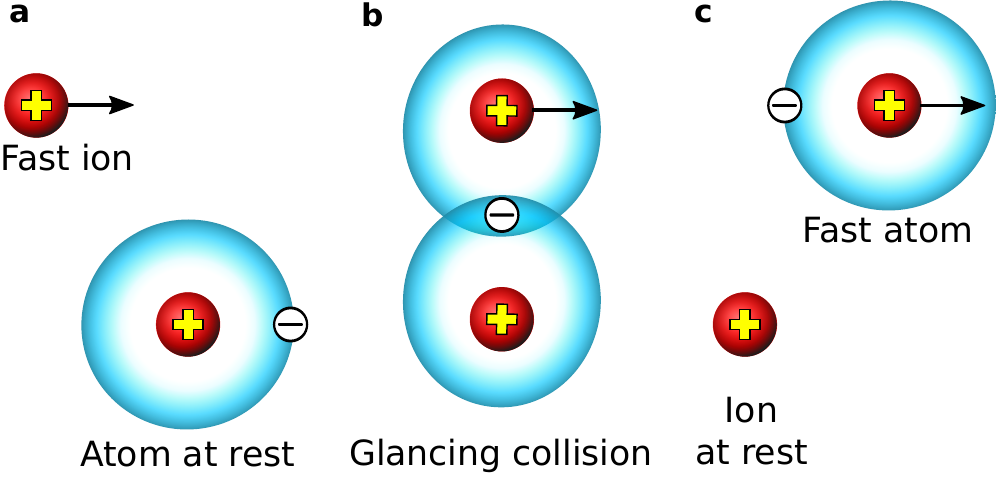}%
	\caption*{\textbf {Fig. 1 \boldmath$\mid$  Scheme for swap cooling process} \textbf{a,} A fast ion approaches a neutral parent atom at rest in a glancing collision. \textbf{b,} When the electron orbitals around the two nuclei overlap, a resonant charge exchange can take place, \blau{because
	the binding energy of the electron is independent of which ionic core it is bound to.} 	
		  \textbf{c,} The new ion is now at rest to the extent that the transferred momentum during the collision is negligible.
	}
	\label{Fig:Scheme}	
\end{figure}
\indent Here we present direct experimental evidence for swap cooling for a single $^{87}$Rb$^{+}$ ion as it passes an ultracold cloud of neutral rubidium atoms. Our scheme allows for a deterministic measurement of the cooling probability of the ion for a given interaction time. We performed our experiments in a regime where sympathetic cooling via elastic collisions is negligible and the
exchange of energy is therefore attributed to swap cooling.
 Moreover, we investigated how swap cooling depends on the initial kinetic energy of the ion and found agreement with theoretical predictions.
Furthermore, in order to highlight the difference between swap cooling and sympathetic cooling, we repeated the cooling experiment with a Ba$^{+}$ ion colliding with  ultracold Rb atoms, a system for which swap cooling mechanism does not exist. As expected, we found that
the highly energetic Ba$^{+}$ ion can only be sympathetically cooled which takes place much more slowly than the swap cooling for Rb$^+$.
We carried out simulations for sympathetic and swap cooling and found good agreement with all our experimental  data.
\begin{figure*}[t!]
	\centering
	\begin{subfigure}[t]{0.36\textwidth}
		\centering
		\caption{}
		\includegraphics[width=\textwidth]{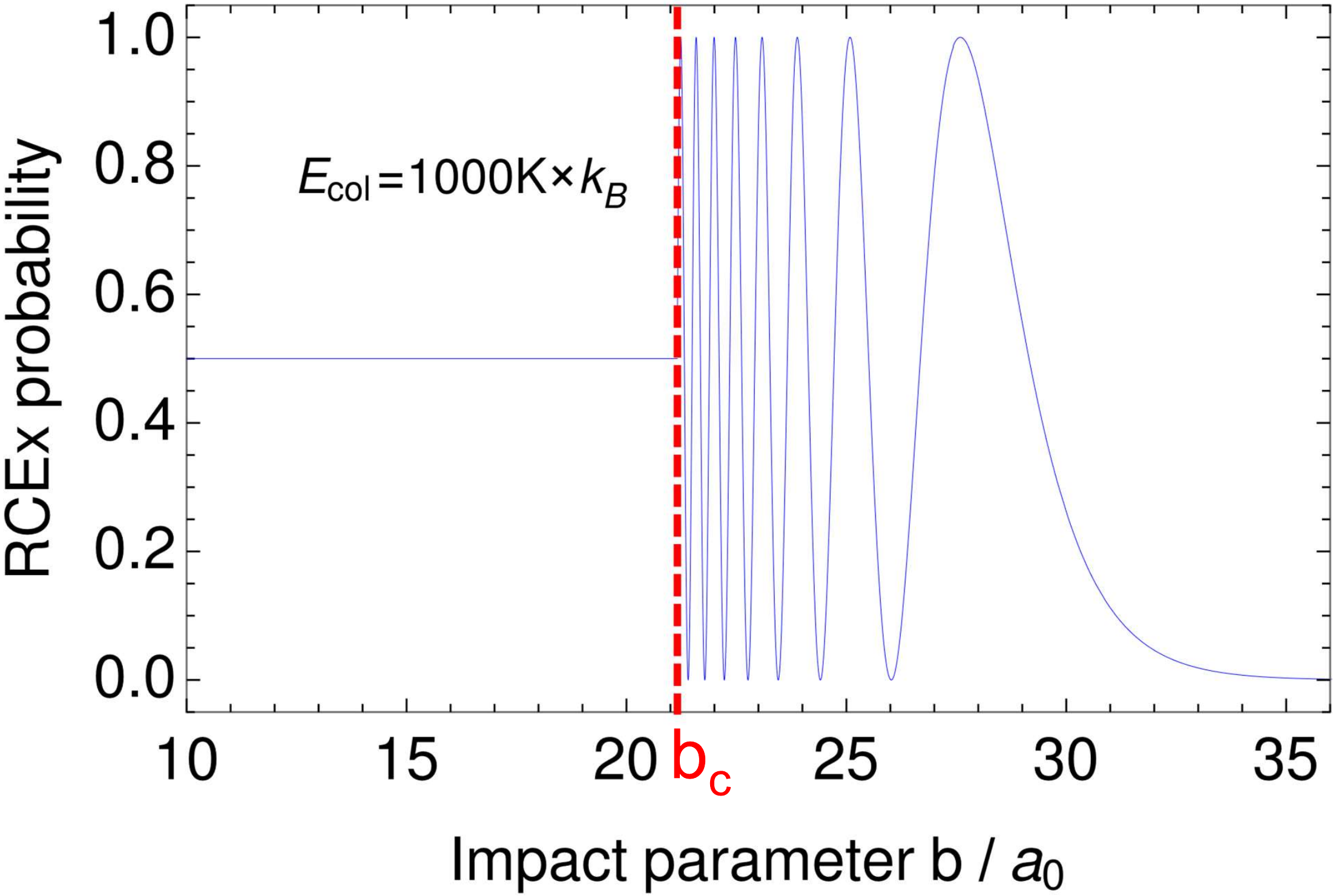}
		\label{fig:RCEx Prob.}
	\end{subfigure}
\hfill
\begin{subfigure}[t]{0.25\textwidth}
	\centering
	\caption{}
	\includegraphics[width=\textwidth]{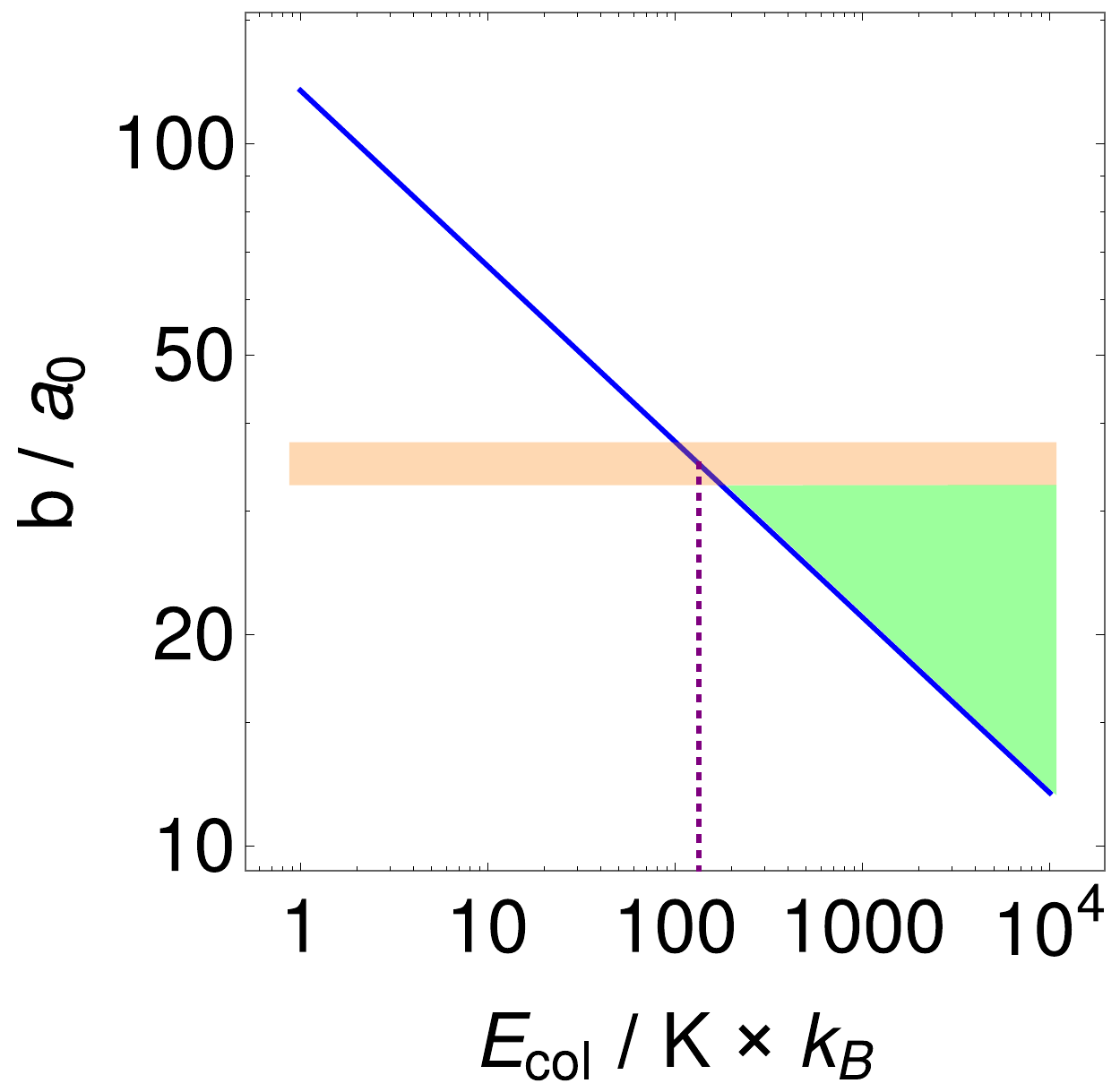}	
	\label{fig:bcVsEcol}
\end{subfigure}
	\hfill
	\begin{subfigure}[t]{0.378\textwidth}
		\centering
		\caption{}
		\includegraphics[width=\textwidth]{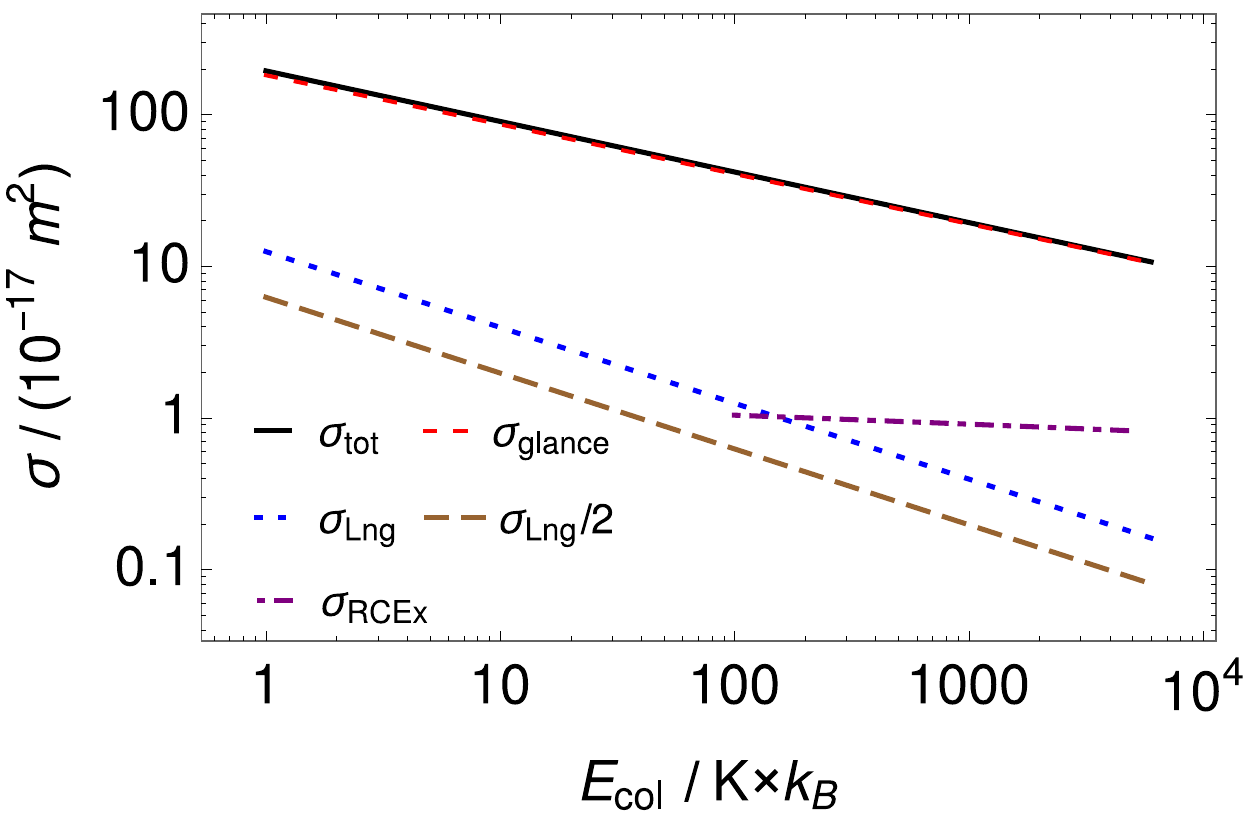}	
		\label{fig:Cross-sections}
	\end{subfigure}
	\caption*{\textbf{Fig. 2 \boldmath$\mid$ Resonant charge exchange, glancing collisions and Langevin collisions. a,} The probability for RCEx in a Rb-Rb$^+$ collision as a function of the impact parameter in a semi-classical picture. The vertical red dashed line marks the critical impact parameter ($b_c$) for the given collision energy $E_{col} = 1000$ K$\times k_B$. Swap cooling occurs when RCEx takes place for  $b > b_c$.	
\blau{The oscillations are a result of the coherent, quantum-mechanical electron hopping between the two ionic cores while atom and ion pass each other, see also text.
		\textbf{b,}
  The solid blue line gives the critical impact parameter  $b_c$ as a function of $E_{col}$.
The horizontal orange line marks the maximum impact parameter for which swap cooling is possible. Thus, for collision energies below $\approx 100$ K$\times k_B$, RCEx can only occur via a Langevin collision (corresponding to $b<b_c$). For $E_{col} > \approx 100$ K$\times k_B$ glancing collisions also contribute in the RCEx. The green area is where swap cooling can take place.
\textbf{c,}
Cross sections for resonant charge exchange $\sigma_{RCEx}$, for Langvin collisions  $\sigma_{Lng}$, for glancing collisions $\sigma_{glance}$, as well as for the total cross section $\sigma_{tot} =  \sigma_{Lng}  + \sigma_{glance} $, as a function of collision energy $E_{col}$.
The Langevin cross section is directly linked to the impact parameter $b_c$ by
$\sigma_{Lng} = \pi b_c^2$ (see also appendix).}
		The RCEx cross section is plotted  only within the range of validity of asymptotic theory (see text). For collision energies below about 100 K$\times k_B$,  $\sigma_{RCEx}$ is essentially given by
		$\sigma_{Lng}/ 2$.
}
	\label{fig:three graphs}
\end{figure*}
\section{Resonant charge exchange collisions}
\indent In general, one can classify atom-ion collisions into two types: Langevin collisions and glancing collisions.
A Langevin collision occurs when the impact parameter $b$ is smaller than a critical value $b_c$ which is
a function of the collision energy $E_{col}$.
In a classical picture, atom and ion fall onto each other in a spiraling motion and finally scatter from the inner-core repulsive potential with an isotropic angle distribution in the center of mass frame.
A glancing collision has $b>b_c$, the atom-ion distance remains on the order of $b$ and generally only leads to small deflections.
Resonant charge exchange (RCEx) can occur in both Langevin and glancing collisions.
Figure 2a shows the probability for resonant charge exchange for a Rb$^+$- Rb collision with a collision energy of 1000 K$\times k_B$ in a semi-classical picture.
The vertical red dashed line marks the critical impact parameter ($b_c$) for  that particular collision energy $E_{col}$, defined in the center of mass frame. For Langevin collisions, $b<b_c$, the electron has an equal chance to end up in each of the two identical (ionic) cores after the collision, which means that there is a 50 percent probability for charge transfer. For glancing collisions the asymptotic theory by Smirnov\cite{15} is used to calculate the RCEx probability (see Appendix). For the given example of Rb, resonant charge transfer can  occur for impact parameters of up to about 35 $a_0$ where $a_0$ is the Bohr radius, leading to swap cooling.
\blau{The oscillations in Figure 2a  can be understood as a result of the coherent, quantum-mechanical electron hopping between the two ionic cores while atom and ion fly by each other. This process is analogous to coherent Rabi-flopping in a two-level quantum system where the coupling is switched on and off in a non-adiabatic fashion. }\\
\blau{  Figure 2b shows the critical impact parameter $b_c$ as a function of the collision energy (blue line). The orange horizontal line shows the maximum impact parameter for which swap cooling is possible.  \blau{In Figure 1a this maximum impact parameter is clearly located at around $b = 30 a_0$.}
Collisions with $b$ below this line but above $b_c$ (blue line) can lead to swap cooling,
 see green area in Figure 2b. The parameter range that contributes to swap cooling clearly increases with $E_{col}$.}

\indent Figure 2c shows calculated cross sections for resonant charge exchange $\sigma_{RCEx}$, for Langevin collisions  $\sigma_{Lng}$, for the total cross section $\sigma_{tot}$, as well as for glancing collisions $\sigma_{glance}=\sigma_{tot} -  \sigma_{Lng}$.
Mathematical expressions for these cross sections can be found in the appendix.
The cross section for RCEx is calculated by means of Smirnov's asymptotic theory which becomes less accurate at lower collision energies\cite{15} and is, therefore,  only shown for energies down to $100$ K$\times k_B$. In fact, for lower collision energies, the critical impact parameter is $>35 a_0$, so that the charge transfer takes place only by Langevin collisions.
Since in a Langevin collision the probability for RCEx is $1/2$,  the charge exchange cross section goes over to $\sigma_{RCEx} = \sigma_{Lng}/2 $  for energies $E_{col} \approx 100$ K$\times k_B$ or below. We further note that since the scattering angle in a Langevin collision is distributed randomly, it is futile to distinguish between an elastic and a charge exchange process in Langevin collisions.
Figure 2c shows that the cross section for glancing collisions $\sigma_{glance}$ is much larger than the cross sections for Langevin collisions or for RCEx.
Nevertheless, the only significant contribution of glancing collisions in collisional cooling is in the form of swap cooling as glancing collisions generally involve very small momentum and energy transfer. Summarizing the results of Fig. 2c, for collision energies in the range from several mK$\times k_B$	to about 100K$\times k_B$ the cooling process is dominated by Langevin collisions.
For larger energies, swap cooling increasingly dominates the cooling process.
 \begin{figure*}
	\centering
\begin{subfigure}[t]{0.495\textwidth}
		\centering
		\caption{}
		\includegraphics[width=\textwidth]{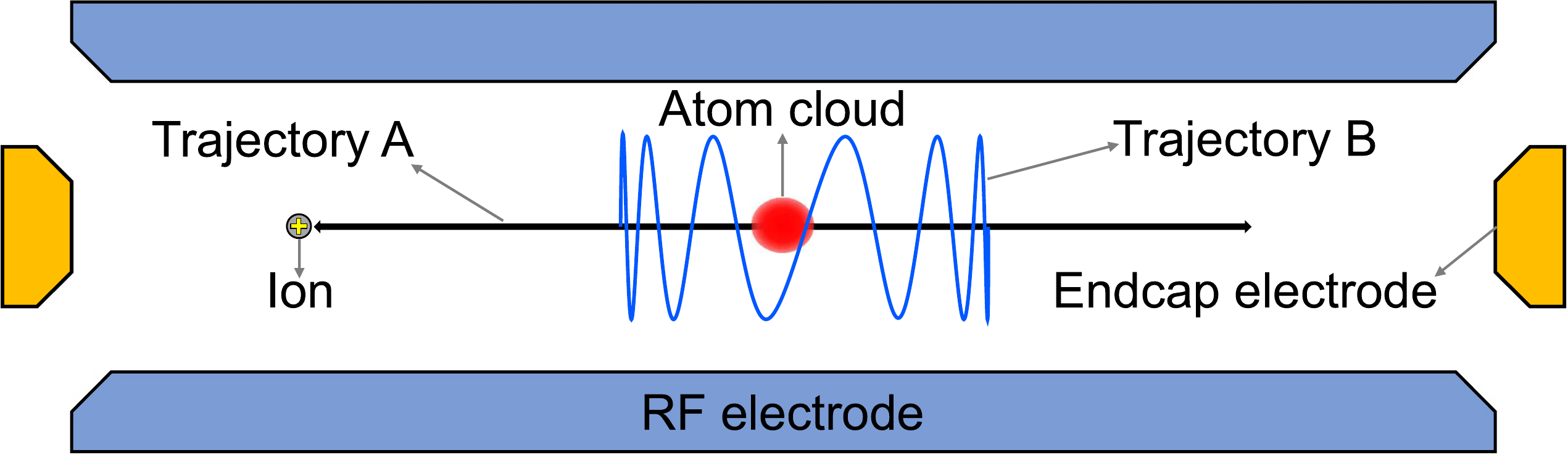}	
		\label{fig:Exp_scheme}
	\end{subfigure}
\hfill
	\begin{subfigure}[t]{0.485\textwidth}
		\centering
		\caption{}
		\includegraphics[width=\textwidth]{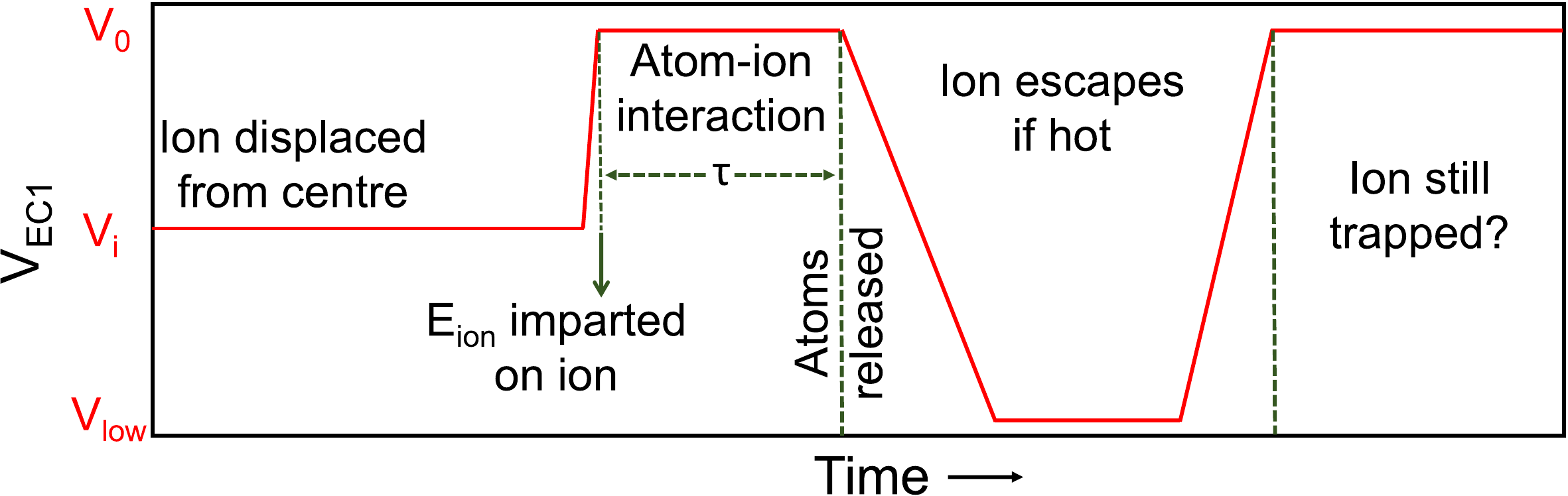}	
		\label{fig:Time_Seq}
	\end{subfigure}
\hfill
	\begin{subfigure}[t]{0.49\textwidth}
		\centering
		\caption{}
		\includegraphics[width=\textwidth]{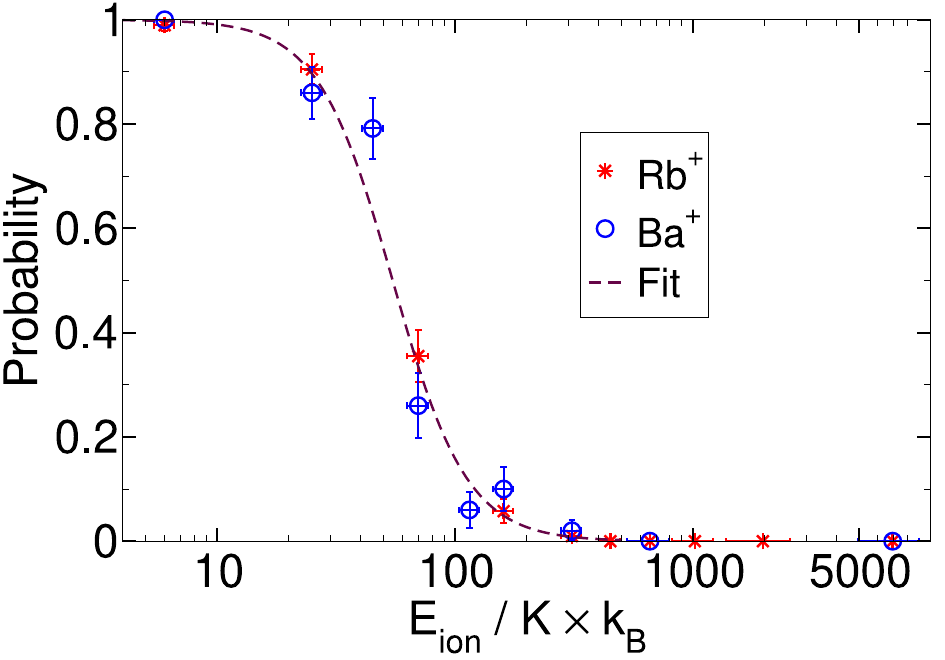}
		\label{fig:Ref}
	\end{subfigure}
	\hfill
	\begin{subfigure}[t]{0.49\textwidth}
		\centering
		\caption{}
		\includegraphics[width=\textwidth]{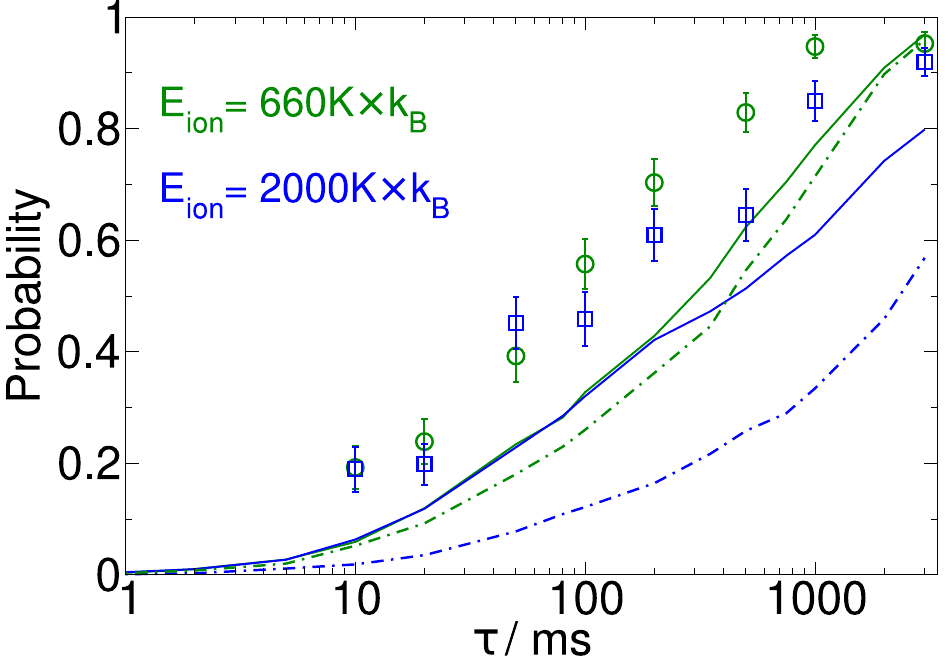}	
		\label{fig:Time_evo.}
	\end{subfigure}
	
	\caption*{\textbf{Fig. 3 \boldmath$\mid$ \blau{Set-up and experiment}}
\textbf{a,} \blau{Sketch of the experimental set-up showing Paul trap electrodes, the ion and the atom cloud.
Two possible trajectories of the ion in the ion trap are displayed. The black, straight-line trajectory represents the initial ion motion along the longitudinal axis ($z$-axis) of the ion trap, after the ion's launch. It runs right through the cloud center. The blue \blau{oscillating} curve is an ion trajectory
	    after a Langevin collision with a cold atom that imparted on the ion a sizable momentum in the transverse direction of the ion trap.	As a consequence, the overlap of the new trajectory with the atom cloud is (on average) strongly reduced. This can significantly slow down further cooling of the ion.}
	    \blau{ \textbf{b,}  Time sequence of experiment. The red solid line is the
voltage $V_\mathrm{EC1}$ on one of the two endcap electrodes.
$V_\mathrm{EC1} = V_i$ corresponds to the initial displacement of the ion.
  For $V_\mathrm{EC1} = V_0$ the ion trap is centered on the
atom cloud. After an interaction time $\tau$ the atom cloud is released and $V_\mathrm{EC1}$ is lowered to $V_\mathrm{low}$, so that a hot ion escapes. Afterwards the voltage $V_\mathrm{EC1}$ is ramped up and we probe whether the ion is still trapped. }
			\textbf{c,} Probability for a Ba$^+$ or a Rb$^+$ ion with energy $E_{ion}$ to remain trapped after the ion trap depth in axial direction has been slowly lowered to a fixed value (see text for details).
		The dashed line is a fit to the experimental results (see text).
		\textbf{d,} Probability for a Rb$^+$ ion to be cooled to an energy below $\approx$50 K$\times k_B$ as a function of the interaction time $\tau$. Plot symbols (circle, squares) are experimental data for initial ion energies (660 K$\times k_B$, 2000 K$\times k_B$), respectively.
		Each data point is extracted from 120 experimental runs.
		Solid lines are full  MC simulations. Dash-dotted lines are MC simulations without charge transfer.
	}
	\label{fig:4}
\end{figure*}
\section{Experiments and results}
\blau{In order to investigate the cooling process, the following experimental procedure was implemented.
In brief, we first accelerated a single ion in a Paul trap towards a cold atom cloud, imparting a well defined kinetic energy to the ion. A sketch of the setup is shown in Fig. 3a). This was done by quickly changing the endcap voltages of the Paul trap. In a second step, after the ion had interacted with the atom cloud,  we determined whether the ion's energy had cooled down below 50 K$\times k_B$.  For this,  we lowered the voltage of an endcap electrode of the Paul trap to an appropriate value, and detected whether the ion was still trapped (The time sequence is shown in Fig. 3b).}
	
In the following, we describe these two steps in more detail.
For initially preparing the ion,  we either loaded a single $^{138}$Ba$^{+}$ ion or a single $^{87}$Rb$^{+}$ ion into a linear Paul trap (see Appendix) and displaced it from the trap center by a variable, but well-defined, distance in the axial direction.
We then loaded an ultracold cloud of $^{87}$Rb atoms into a crossed optical  dipole trap which was positioned at the center of the Paul trap. The cloud size was smaller than the displacement of the ion, such that initially the ion could not interact with the atoms.
\blau{After the initial displacement of the ion, it was accelerated} towards the atoms by suddenly moving the center of the trapping potential back to the middle of the atom cloud \blau{(see Fig. 2b and Appendix)}.  As a consequence the ion was oscillating within the Paul trap  along the axial direction,  periodically crossing the cold atom cloud.
The kinetic energy of the ion $(E_{ion})$ at the location of the atoms was initially precisely defined by the ion's initial displacement and by how fast the trap center was moved back. $E_{ion}$ could be tuned over several orders of magnitude, up to about 7000 K$\times k_B$. After a time $\tau$ during which the ion could collide with the atoms, we determined whether the ion had a energy
(in axial direction) below $\approx$50 K$\times k_B$ and therefore learned whether it had been collisionally cooled below this threshold.
For this, we removed the atoms and slowly lowered the depth of the ion's trapping potential in axial direction to a fixed value (see Appendix for the details), such that the ion would only stay in the trap
and could be detected if its energy had been cooled to below $\approx$50 K$\times k_B$.
 From many repetitions of the measurement we obtained the probability for cooling below this (somewhat soft [as explained below]) energy threshold within $\tau$. In order to test and calibrate this method
we carried out reference measurements where we probed an ion with a known initial energy  $E_{ion}$. For this, all experimental settings were the same, except that there was no atom cloud present. Figure 3c shows these measurements for both Ba$^+$ and Rb$^+$.
The dashed line is a phenomenological fit of the form
$y=1 / ((E_{ion}/a)^{b}+1)$
to the experimental data, where $a = 54$ K$\times k_B$ and $b = 2.76$.
The fact that the curve is not a step function but has 80\%-20\% width of about 60 K$\times k_B$ is mainly due to the fact that the turn-down of the trap potential is not fully adiabatic and that it is not synchronized with the ions oscillatory motion.
Nevertheless, Fig. 3c shows that once the ion has an axial energy lower than 50 K$\times k_B$, it will have a high probability to stay in the trap. \\
\indent We now describe the cooling experiments. Figure 3d shows the probability to cool
a highly energetic Rb$^+$ ion as a function of the interaction time
$\tau$. We show data of two different experiments, corresponding to an initial kinetic energy of the ion of  (660$\pm$65)  K$\times k_B$, (green circles) and (2000$\pm$300) K$\times k_B$ (blue squares). As expected, the cooling probability monotonically increases with $\tau$. However, what might be surprising at first is that the cooling results for the two experiments are very similar, although their initial
kinetic energies differ by almost a factor of three. This cannot be explained within standard sympathetic cooling - but is an indication of swap cooling which can cool an ion in a single step. Another curiosity is that cooling seems to slow down as time $\tau$ increases. Already at $\tau = 10$ ms about 20\% of the ions have been cooled down. At that cooling rate one might expect to reach a cooled fraction of nearly 100\% by about 150ms. However, in the experiments we only get close to  this limit after about 2 s. The main explanation for this behavior is as follows: If in a collision (e.g. a Langevin collision) the ion obtains a sizable momentum kick in the radial direction, then there is a good chance that its new trajectory has a much smaller overlap with the atom cloud than before (see Fig. 3a). This drastically lowers the subsequent cooling rate.
These scenarios are corroborated by Monte Carlo (MC) trajectory calculations for our experimental settings (solid and dash-dotted lines in Fig. 3d). Details of the model are given in the Appendix. The solid lines in Fig. 3d are the full model, whereas the dash-dotted lines are a model without the RCEx process, i.e., without swap cooling. The model without swap cooling is inconsistent with the data as it clearly distinguishes between the two cases with different initial energies and predicts a much slower cooling rate, especially for the higher energy. The full model reproduces the overall behavior of the measured data but seems to underestimate the cooling rate.  This might be partially explained by systematic errors in the theory when applying it to our energy range. The theory by Smirnov \cite{15} is expected to be more accurate for keV collision energies and assumes the particle trajectories to be straight lines, which is probably not justified for our experiments.

\indent
In order to obtain more evidence for swap cooling, we determined the energy dependency of the cooling process. Furthermore, we compared cooling of a single $^{87}$Rb$^{+}$ ion with that of a single $^{138}$Ba$^{+}$ ion,
for which RCEx, and therefore swap cooling, is absent.
Figure 4 shows the probabilities for cooling each ion with an initial energy\footnote{For Rb$^+$ $E_{ion} = 2\times E_{col}$. For Ba$^+$, assuming the Rb atom is at rest, $E_{ion} = m_{Ba}/ m_{Red} \times E_{col} = 2.59 E_{col}$ where $m_{Red}=\frac{m_{Ba}m_{Rb}}{m_{Ba}+m_{Rb}}$ is the reduced mass in the two-body collision.} $E_{ion}$ to below $\approx$50 K$\times k_B$ within $200\; $ms. For low initial energies of up to 200 K$\times k_B$ both ions
are essentially completely cooled down within the 200 ms interaction time.  This is not surprising as the dynamics of the system is described by the Langevin collisions in this range of energy.
By increasing the initial energy $E_{ion}$, the probability for cooling below the threshold decreases for both species, as expected. However, the cooling efficiency differs progressively  for the two species as $E_{ion}$ gets larger. For Ba$^+$ the cooling probability essentially drops to zero for $E_{ion}=7000 $ K$\times k_B$. This is simply explained by the fact that swap cooling does not exist for Ba$^+$ and the time $\tau$ is too short for sympathetic cooling to reach the $\approx 50$ K temperature threshold. For Rb$^+$, however, the probability to cool down the ion is still above 40\%.
The solid red line shows the cooling results for full MC calculation for Rb$^+$. The red dotted line is the MC calculation for Rb$^+$ without RCEx but including sympathetic cooling via elastic collisions. The blue solid line shows the MC simulation results for sympathetic cooling of the Ba$^+$ ion. As expected, the last two are very similar apart from a small horizontal shift.
The solid curves match the general trend of the experimental data quite well, which is consistent with our previous discussion.
\begin{figure}
	\centering	
	\includegraphics[width=0.46\textwidth]{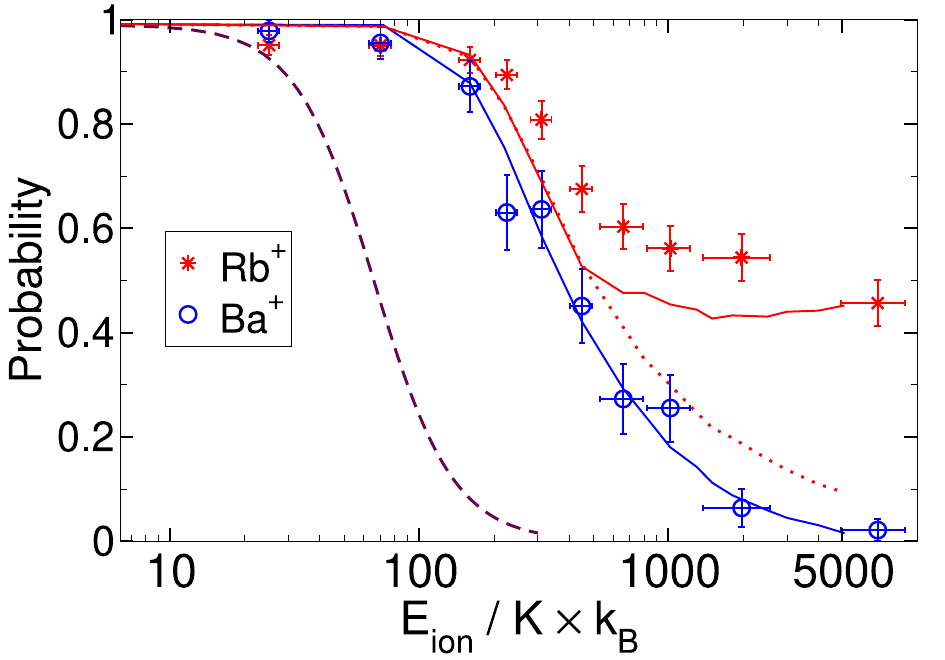}
	\caption*{\textbf{Fig. 4 \boldmath$\mid$  Cooling of Rb$^\textbf{+}$ and Ba$^\textbf+$ ions.}
		Shown is the probability to cool an ion to below $\approx 50$ K within $\tau = 200$ ms with ultracold cloud of Rb atoms. The probability is plotted as a function of the ion's initial kinetic energy $E_{ion}$.	
		The dashed line is the reference curve from Fig 3a. The solid lines are full MC simulations.  The dotted red line is MC simulation without RCEx for a Rb$^+$ ion.
	}
\end{figure}
\section{Conclusion}
The swap cooling process, demonstrated here for Rubidium, is based on a few very general principles and should therefore be a universal process in all homo-nuclear atom-ion collisions. Furthermore, it should even occur for molecules.
In addition to this proof of principle result, we anticipate the cooling procedure to have
 practical applications, e.g. where ions with energies in the eV \blau{to keV range and beyond need to be cooled down quickly and efficiently.
 The reason that swap cooling should work well even at keV energies and beyond is due to the fact, that according to  Smirnov's model the total cross section for resonant charge exchange only decreases extremely slowly as a function of collision energy.
Besides the high temperature limit, } swap cooling should also play an important role in current cold atom-ion experiments, such as in \cite{13} where  a chemical reaction process was driven by cold Rb$^+$ ions and emitted hot ones. Swap cooling is then an efficient means to keep the reaction process going. Moreover, it is essential to take into account the RCEx in investigating the mobility of a charged particle in parent neutral particles\cite{16}.
\section*{Appendix}
{\bf Preparation of the ultracold Rb clouds.} The atoms are confined in a far off-resonant crossed optical dipole trap at a wavelength of 1064.5 nm with a trap depth of 10 $\mu$K$\times k_B$ and with trap frequencies of $\omega_{x,y,z}=2\pi\times(24,138,145)$ Hz. Atom clouds are prepared differently depending on what they are used for. Atom clouds that are used as coolant or as a means for ion detection typically consist of $N=7\times10^4$ atoms. They are thermal ensembles with  temperatures of about $T=600$ nK. Their cloud shape is Gaussian with a standard deviation of $\sigma_{x,y,z}\approx(48, 8.8, 8.4)$ $\mu$m, and with a peak density of $n=1.3\times10^{12}$ cm$^{-3}$.
Atom clouds that are used for producing the Rb$^+$ ions (see section `Preparation of single ions at mK-temperatures') have a total atom number $N=1.2\times10^6$ corresponding to peak densities higher than $10^{13}$ cm$^{-3}$. The initial temperature of these clouds is about $T=510$ nK.\\

{\bf Ion trap configuration.}
We use a linear Paul trap\cite{17} where radial confinement is achieved  by driving the RF-electrodes with a radio-frequency of $4.21\;$MHz and axial
confinement is obtained by applying static voltages of $7.6$ V to the two end cap electrodes.
The trap frequencies of a Ba$^+$ ion are $(\omega_{x,Ba},\omega_{y,Ba}, \omega_{z,Ba}) = 2\pi\times(130,131,37.8)\;$kHz, while for a Rb$^+$ ion they are $(\omega_{x,Rb},\omega_{y,Rb}, \omega_{z,Rb}) =2\pi \times (206,208,47.6)$ kHz.
The depth of the Paul trap depends on the ion's mass but exceeds 1 eV for both ions.\\

{\bf Preparation of single ions at mK-temperatures.} We prepare a single mK-cold  Rb$^+$ ion as follows. We produce an ultracold Rb cloud with a density of $n > 10^{13}$ cm$^{-3}$ in our dipole trap. Three-body recombination  creates Rb$_2$ molecules within this cloud, a few of which are photo-ionized by our dipole trap laser in a resonant multi-photon ionization (REMPI) process\cite{18}.
In this way we produce Rb$_2^{+}$ molecular ions at an initial rate of about 0.5 s$^{-1}$. Subsequent inelastic collisions of these ionic molecules with cold Rb atoms, or photodissociation via the 1064 nm laser light produce Rb$^{+}$ ions. The Rb$^+$ ion will then have elastic collisions with Rb atoms  which will decrease the density of the atoms by heating the cloud and kicking atoms out of the shallow trap. Therefore, the production rate of additional ions via three-body recombination quickly drops after the production of the first ion. As a consequence, with the proper settings of the experimental parameters we can prepare a single Rb$^{+}$ ion with more than 90 percent probability.
In order to experimentally test whether we have  created a single Rb$^{+}$ ion and to sympathetically cool it down to mK temperatures,  we  immerse the ion/ions for two seconds  into a newly prepared Rb atom cloud with a comparatively low density of about $10^{12}$ cm$^{-3}$. We make use of ion-induced atom loss, in the same manner as explained in section `Detection of ions', in order to determine the number of the ions in the trap. For our analysis, we only take into account experimental runs which initially had a single ion.
If we need to prepare a single cold Ba$^+$ ion, instead of a Rb$^+$ ion, we use isotope-selective, resonant two-photon ionization of neutral Ba atoms passing through the center of the Paul trap. Once produced, a Ba$^+$ ion is immediately trapped  and laser cooled down to the Doppler cooling limit.\\

{\bf Setting the initial kinetic energy of the ion.}
For preparing the ion with desired initial kinetic energy after it has been cooled to mK temperatures, the center of the axial trapping potential of the Paul trap is adiabatically shifted by slowly lowering the voltage of one of the end cap electrodes which leads to an adiabatic displacement of the ion. Afterwards, the voltage of the end cap is set to its original value using a well-defined voltage ramp.
This imparts a well defined kinetic energy to the ion as it swings back to the
trap center where the atom cloud is located. The ion then undergoes essentially undamped harmonic oscillations in the trap along the axial direction until the first collision with a cold Rb atom which can change the ion momentum appreciably.\\
\indent
In order to calibrate the ion's energy, we measure the initial displacement of the ion as a function of the applied voltages to the end cap electrodes. The mean displacement of the ion can be measured with an accuracy on the order of 1 percent via fluorescence imaging of the Ba${^+}$ ion. Since the minimum of the trapping potential is independent of the mass, the displacement is identical for Rb${^+}$. The imparted energy to the ion is determined by solving the equation of motion numerically, making use of the known axial trapping potential and including a 30 V/m stray electric field. As a typical  example, linearly ramping the voltage on the
end cap electrode from 0.6 V to 7.6 V in 9.4 $\mu$s imparts an energy of 2000$\pm$ 300 K$\times k_B$ to the ion.\\

{\bf Determining whether ions have been cooled down to 50 K$\times k_B$ or below.} We use a scheme which is rather simple and empirical but gives a clear semi-quantitative measure of whether the ion energy has been cooled down in axial direction  to $\approx 50$ K$\times k_B$ or below by the atom cloud. Right after the atom-ion interaction, we lower the voltages on the electrodes of the Paul trap and hold  them at that level for 200 ms. The ramp-down in the axial direction is done by slowly decreasing the voltage on one of the
end cap electrodes (from 7.6 V to around 100 mV) within 150 ms.
We note that this process also shifts the trap center in the axial direction, which in general leads to some heating of the ion.
Simultaneously, we lower the RF amplitude in the radial direction from 156 V to 50 V within 150 ms.
These experimentally determined values for the voltages on the Paul trap
are the lowest ones that allow mK-cold ions to still  be trapped with unit probability after ramping.
We  measure the probability of losing the ion after the lowering process as a function of its initial kinetic energy in the axial direction. These calibration measurements are shown in Fig. 3a. Losses in the radial direction should be negligible because the radial confinement is still comparatively large. Moreover, our simulations for large initial energies $>$ 1000 K$\times$k$_B$ indicate that after the 200 ms interaction time of our measurements, the ion's kinetic energy for the axial motion is typically still much larger than the one for the radial motion. Therefore, in this energy range the measured energy in axial direction should be quite representative of the overall energy of the ion.\\

{\bf Detection of ions.}
In order to detect how many ions are present in the Paul trap, we immerse them for
two seconds into a newly prepared Rb atom cloud with a density of about $10^{12}$ cm$^{-3}$ and with a well-defined atom number of $(7\pm 0.2)\times 10^4$. Due to their micromotion the ions induce loss in the atomic cloud as a result of elastic collisions which kick atoms out of the shallow dipole trap. We infer the ion number from a measurement of the atom loss \cite{13}. \blau{However, in order to  specifically detect a single Ba$^+$ ion  we use a different method, where we perform intensive laser cooling by detuning the laser frequency 2.5 GHz to the red of the resonance and ramping it back linearly in about 4s \cite{10b} We have checked that this procedure  cools down even the hottest Ba$^+$ ions (which have a large Doppler shift) in our Paul trap with nearly 100 percent efficiency.} Once the Ba$^+$ ion is cooled down we collect its fluorescence signal on a EMCCD camera within 100 ms.\\

{\bf Theoretical model for swap cooling.}
In the following we describe the semi-classical model for swap cooling in a homo-nuclear atom-ion collision.
The long range atom-ion interaction is dominated by the attractive polarization potential\cite{19}
$V(r)=-C_4/(2r^4)$, where $C_4=\alpha q^2 / (4\pi\epsilon_0)^2$
is proportional to the static dipolar polarizability of the atom $\alpha$, and $q$ is the electron charge.  $\alpha = 4\pi \epsilon_0 \times 47.4$ \AA$^3$ for Rb\cite{33}. For a given collision energy $(E_{col})$ in the center of mass (COM) frame, a critical impact parameter  $b_c = (2C_4/E_{col})^{1/4}$ can be defined. In a classical picture, a collision with impact parameter $b<b_c$ results in an inward spiraling trajectory, followed by a hard elastic or inelastic collision, and completed by an outward spiraling trajectory. The scattering angles for such a collision are uniformly distributed over the whole solid angle of $4\pi$. This is called a Langevin collision and its cross section is $\sigma_{Lng}=\pi\, b_c^2 =\pi \sqrt{{2C_4}/{E_{col}}}$. Collisions with impact parameters $b>b_c$ generally lead to small deflections and are therefore called glancing collisions.
They are of interest for the swap cooling process.
For glancing collisions, the atom-ion separation stays finite and the two particles are scattered from each other with a deflection angle $\theta$ which is a function of the impact parameter $b$. For a $1/r^4$ potential the angle
$\theta$ is given by \cite{20},
\begin{equation}
\theta=\pi - 2\tilde{b}\sqrt{2}\sqrt{\tilde{b}^2-\sqrt{\tilde{b}^4-1}}\; \times \;\mathcal{K}\;(2\tilde{b}^4-2\tilde{b}^2\sqrt{\tilde{b}^4-1}-1)
\label{eq:theta_equation}
\end{equation}
where $\tilde{b}=b/b_c$ is the normalized impact parameter and
$$
\mathcal{K}(y) = \int_{0}^{\pi/2}\frac{1}{\sqrt{1-y\sin^2x}}dx
\label{eq:K_y}
$$
is the complete elliptic integral of the first kind.
In the following we consider the collision of an incident Rb$^+$ ion with a neutral Rb atom which is at rest in the lab frame. For an elastic collision, the energy $E_{loss}$ lost by the incident particle in the lab frame is equal to the energy $E_{trans}$ transferred to the target particle which is initially at rest. These energies
can be expressed as a function of the scattering angle $\theta$ in the COM frame (see e.g. \cite{20}),
\begin{equation}
E_{loss}= E_{trans} = E_{ion,i}\sin^2({\theta/2}),
\label{eq:E_loss}
\end{equation}
where $E_{ion,i}$ is the initial kinetic energy of incident ion.
It is noteworthy that eq. (\ref{eq:E_loss}) is merely a consequence of the conservation of momentum and energy.
We now discuss the probability for resonant-charge-exchange in a collision.
RCEx has been  extensively studied theoretically\cite{21,22,23} and has been experimentally investigated for  collision energies ranging from eV to keV for different species, including Rubidium\cite{24}. Here we use the impact parameter dependent charge exchange probability $P_{RCEx}(b)$ obtained from the asymptotic theory  presented by Smirnov\cite{15} which in atomic units reads,
\begin{equation}
\begin{split}
&P_{RCEx}(b)=\\
&\sin^2\left( \frac{1}{v}\,\sqrt{\frac{\pi}{2\gamma}}\, A^2 \,\exp(-\frac{1}{\gamma})\,b^{2/\gamma-1/2}\,\exp(-b\gamma) \right).
\end{split}
\label{eq:P_RCEx}
\end{equation}
Here, $v$ is the collision velocity (1 a.u. of velocity $ = a_0 E_h/ \hbar = 2.187 691 263 64 \times 10^6$ m s$^{-1}$), $\gamma = \sqrt{- 2 I}$ where I is the atomic ionization potential, $A$ is the asymptotic coefficient determined by comparing the asymptotic wave function with that obtained from numerical calculations, and $b$ is the impact parameter in the collision. We adopt the numerical values of $\gamma = 0.554$ and $A = 0.48$ for Rubidium from reference\cite{15}.
The  total cross section $\sigma_{RCEx}$ for resonant charge exchange is then,
\begin{equation}
\quad \sigma_{RCEx}=\int_{0}^{\infty}2\pi\, b\, db\, P_{RCEx}(b)
\label{eq:sigma_RCEx}
\end{equation}
For swap cooling we are only interested in those charge transfer collisions that result in an ion which is cold enough. Therefore, we define the following swap cooling cross section,
\begin{equation}
\sigma_{swap}(b_{thr}) =\int_{b_{thr}}^{\infty} 2\pi\, b\,db\,P_{RCEx}(b).
\label{eq:sigma_swap}
\end{equation}
This cross section depends on the threshold impact parameter $b_{thr}>b_c$.
Every charge transfer collision with an impact parameter greater than $b_{thr}$ produces a Rb$^+$ ion with a kinetic energy less than a maximal final energy $C_{thr}= E_{ion,i}\sin^2({\theta (b_{thr})/2})$ which follows from eqs. (\ref{eq:theta_equation}) and (\ref{eq:E_loss}). All these collisions are accounted for in the swap cooling cross section (\ref{eq:sigma_swap}).
The calculated results for the swap cooling cross section are presented in Figure 5 for different maximal final energies $C_{thr}$.  We also show the total resonant-charge-exchange cross section from eq. (\ref{eq:sigma_RCEx}) \blau{which is decreasing only extremely slowly with collision energy. It is worth noting that the swap cooling cross sections increase by increasing the collision energy, approaching the total resonant-charge-exchange cross section.
The increase is more pronounced for lower maximal energies ($C_{thr}$) of the ion after the collision.
The extremely slow decrease of the total resonant-charge-exchange cross section with collision energy
suggests that swap cooling should work very well even at much higher energies than the $\approx$ 7000 K$\times k_B$ range investigated experimentally in our work.  Indeed, judging from Smirnov's theory we expect it to work up to the keV range and beyond.
}
\\
\begin{figure}
	\centering	
	\includegraphics[width=8cm]{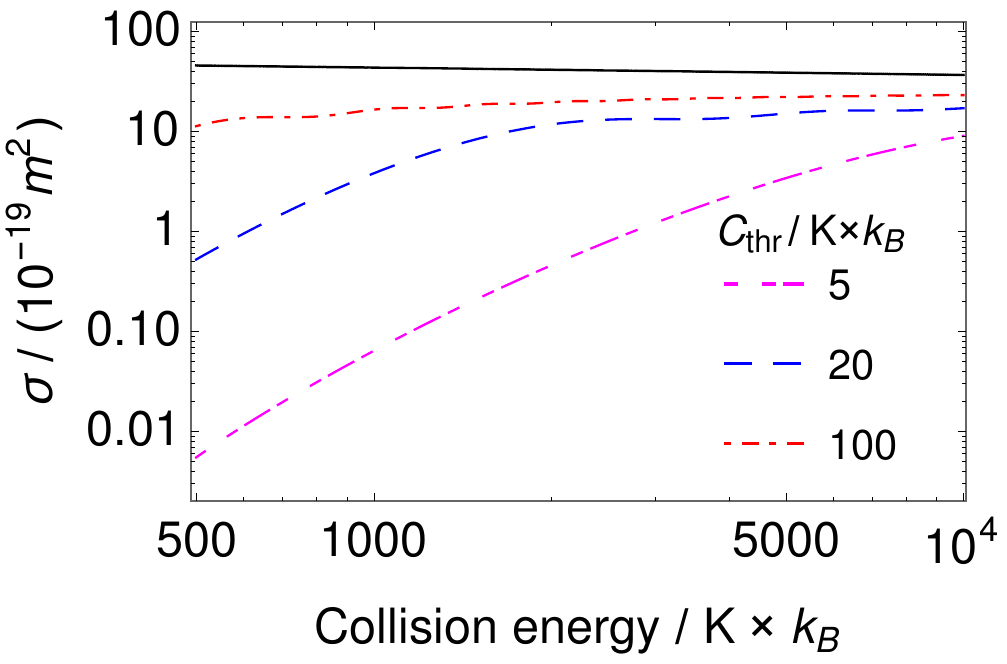}
	\caption*{\textbf{Fig. 5 \boldmath$\mid$ The RCEx and swap cooling cross sections for a Rb-Rb$^+$ collision.} The black curve is the calculated RCEx cross section using asymptotic theory\cite{15}.
		Three different swap cooling cross sections are also presented for different maximal energies ($C_{thr}$) of the ion  after the collision.}
\end{figure}\\

{\bf Description of the simulation.} We have performed Monte Carlo trajectory calculations for the motion of the ion in the ion trap and its collisions with the ultracold Rb atoms. These collisions  include Langevin collisions,
resonant charge exchange collisions and elastic collisions.
The simulation takes into account the settings of the experiment, such as the ion trap frequencies, the density distribution of the atomic cloud, and the ion's initial position and energy.
We use the semi classical elastic scattering cross section of reference \cite{16}
as a total cross section $\sigma_{tot}$ for both elastic and charge transfer processes,
\begin{equation}
	\sigma_{tot}(E_{col})=\pi\,\bigg (\frac{\mu C_4^2}{\hbar^2}\bigg )^{1/3}\,\bigg(1+\frac{\pi^2}{16}\bigg)\, E_{col}^{-1/3}.
	\label{eq:sigma_SemiClass}
\end{equation}
The finite value of $\sigma_{tot}$ effectively limits the impact parameter to  $b \le b_m = \sqrt{ \sigma_{tot} / \pi }$. This limitation does not affect the results, as the trajectories with larger impact parameters have both negligible deflection angles and energy transfer.
The rate for a collision is given by
\begin{equation}
	\Gamma = n_{atom} \, \sigma_{tot} \, v,
\end{equation}
where $n_{atom}$ is the density of the atom cloud at the position of the ion, and $v$ is the ion's velocity in the lab frame. The ultracold atoms are considered to be at rest.
We follow the method in reference \cite{25} for efficient time advance in the trajectory calculations. Once a collision occurs we generate the impact parameter $b$ via\footnote{This formula corresponds to  the inverse of the integrated probability distribution $P(b) = \int_0^b dP =  \int_0^b 2\pi b' db' /\sigma_{tot} = (b/b_m)^2 $} $b = b_m \sqrt{N_{Rand}} $ where $N_{Rand}$ is
a random number drawn from a uniform distribution in the interval $[0 \dots 1]$. The impact parameter determines whether the collision is of the Langevin type or a glancing collision.
For a Langevin collision ($b<b_c$) the scattering angle $\theta$ in the center of mass system is isotropically distributed. For the glancing  collisions ($b_m>b>b_c$) which include both elastic and RCEx processes, the scattering angle $\theta$ is determined by eq.~\ref{eq:theta_equation} and the transferred energy to the target particle (in the lab frame) is given by eq.~\ref{eq:E_loss}.
The probability for RCEx is calculated via eq.~\ref{eq:P_RCEx}. The simulation runs until the total interaction time
is reached. If the final energy of the ion in axial direction is smaller than a set threshold around
$50$K$\times k_B$, it will be counted as a cold ion which stays trapped in the Paul trap after the ramp down.
In order to reproduce the finite energy resolution of the detection scheme (as shown in Fig. 3a) we generate  a corresponding distribution of random numbers for the  threshold in our simulation.  In order to finally calculate the probability for the ion to stay trapped for a given threshold, we analyze typically 10000 MC runs and average over the trapping probability for all thresholds of the distribution.

During the Monte Carlo simulation the atom cloud density distribution is assumed to be constant. This is reasonable since atomic losses due to background collisions, heating and atom-ion collisions during the interaction time are negligible.

\section*{Acknowledgments}
AM appreciates generous support from the European Commission under the Seventh Framework Programme FP7 GA 607491 COMIQ, as well as financial support from the SFB/TR21 graduate school program and from a DAAD Fellowship. The authors acknowledge further support through project A4 of SFB/TR21 of the  Deutsche Forschungsgemeinschaft (DFG).  The authors appreciate helpful
support from J. Wolf, A. Mohammadi, and M. Dei\ss. JHD would like to thank Sadiq Rangwala for interesting discussions.

\section*{References}

\footnotesize \renewcommand{\refname}{\vspace*{-30pt}}
\bibliographystyle{ieeetr} 

\end{document}